\documentclass[aps,prl,twocolumn,showpacs,floatfix,superscriptaddress,graphics]{revtex4-1}

\usepackage{amsmath,amssymb,graphicx,color,braket,hyphenat,makeidx,xcolor,dsfont,subfigure,soul}
\usepackage[ocgcolorlinks,colorlinks=true,linkcolor=blue,citecolor=red,linktocpage=true]{hyperref}

\newcommand{\blue}{\color{\blue}}

\definecolor{coolblack}{rgb}{0.0, 0.18, 0.39}

\begin{document}

\title{Comment on ``Steady-State Coherences by Composite System-Bath Interactions''}

\author{Marco Cattaneo}
\affiliation{QTF  Centre  of  Excellence,  University  of  Helsinki,  P.O.  Box  43,  FI-00014  Helsinki,  Finland}
\affiliation{IFISC, Institute for Cross-Disciplinary Physics and Complex Systems (UIB-CSIC), Campus Universitat Illes Balears, E-07122 Palma de Mallorca, Spain}


\author{Gonzalo Manzano}
\affiliation{IFISC, Institute for Cross-Disciplinary Physics and Complex Systems (UIB-CSIC), Campus Universitat Illes Balears, E-07122 Palma de Mallorca, Spain}

\maketitle


In a recent Letter, Guarnieri {\it et al.}~\cite{Guarnieri:2018} reported on the possibility of generating steady-state coherence (SSC) in the energy basis of a two-level system  by the sole action of a single thermal bath
also for weak system-bath coupling and due to a composite form of this interaction.
The results are based on a time-local Bloch-Redfield master equation without secular approximation and the associated set of Bloch equations, all derived in the Supplemental Material (SM) of Ref.~\cite{Guarnieri:2018}.
We comment that the results reported by the authors present both technical flaws and conceptual limitations that substantially impact some of their main conclusions.
Our objections mainly concern (a) technical errors in the calculation of SSC (Eqs.~(2)-(4) in \cite{Guarnieri:2018}) and (b) negativity issues associated with the weak-coupling master equation employed by the authors. 

\begin{figure}
 \includegraphics[width=\linewidth]{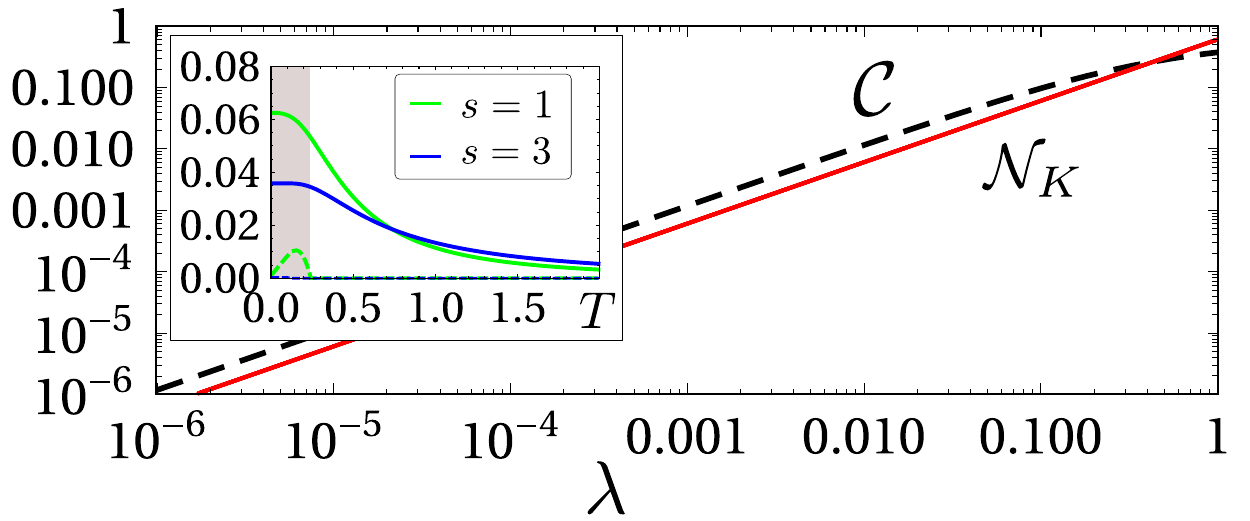}
 \caption{Optimal SSC $\mathcal{C}$ and negativity $\mathcal{N}_K$ as a function of $\lambda$ for Ohmic spectral density, $T = \omega_0$ and other parameters as in Fig.~2(a) of Ref.~\cite{Guarnieri:2018}. Inset: Optimal SSC (solid lines) as a function of temperature $T$ in $\omega_0$ units for Ohmic ($s=1$) and super-Ohmic ($s=3$) cases, together with the corresponding negativities of $\rho_\mathrm{ss}$ (dashed lines), highlighted by the shaded area. Since $\mathcal{C}$ is a growing function of $f_1$ and $f_2$ in the regime $f_1,f_2 \sim O(1)$, without losing generality we used  $f_1=f_2=1$.}
 \label{fig:figure}
\end{figure}

(a) We start by giving the correct expression for the SSC derived from the Bloch-Redfield master equation. We obtain it by correcting a minus sign in the function $\Delta_1(T, \Omega)$ in the first line of Eq.~(3), the sign of the inhomogeneous term in the equation for $\langle \dot{\sigma_y} \rangle$ (term  $b_2(t)$ in Eq.~(32) of the SM), and including the correct expression of the function $\gamma_1(+\infty) =  2 \pi \lim_{\omega\rightarrow 0^+} J_\mathrm{eff}(\omega, T) = 4 \pi \lambda T \delta_{s,1}$ with $s \geq 1$. The latter induces extra dephasing on the two-level system for Ohmic environments $(s=1)$, while $\gamma_1(+\infty)$ diverges in the sub-Ohmic case $(s<1)$, thus leading to zero SSC~\cite{Sabrina:2014}. For the Ohmic case we obtain:
\begin{equation}\label{eq:v1}
 \bar{v}_1 = f_1 f_2 \frac{\Delta_1 \tanh(\frac{\omega_0}{2T})-4 \lambda \Omega \Gamma(1)-\Delta_2}{\omega_0 + f_2^2 \Delta_1 + f_1^2 2 T \lambda \Delta_1/J_\mathrm{eff}(\omega_0,T)} 
\end{equation}
and $\bar{v}_2 = 0$, where we used the same notation as in Ref.~\cite{Guarnieri:2018}, c.f. Eq.~(2) therein.
Note that the new expression displays some adjusted signs and an extra term in the denominator. This severely reduces the resulting SSC, $\mathcal{C} \equiv \sqrt{\bar{v}_1^2 + \bar{v}_2^2} = |\bar{v}_1|$, its optimization w.r.t. $f_1$ and $f_2$, and the curves reported in Fig.~2(a) of~\cite{Guarnieri:2018} c.f. Fig.~\ref{fig:figure} (inset) [similar errors apply to 
Fig.~2(b)]. We also notice that the SSC enhancing ``spykes'', there attributed to the properties of the resonance curve in sub-Ohmic and Ohmic regimes, correspond instead to unphysical divergences [zeros in the denominator of Eq.~(4)]. 
These emerge because the maximization performed in Eq.~(4) does not respect weak-coupling [which requires $f_1,f_2 \sim O(1)$], but similar divergences also arise in Eq.(2) for higher temperatures $k_B T \simeq 30 \omega_0$. 

(b) We now give evidence that the SSC predicted by this model 
are linked to the presence of 
negative eigenvalues in the so-called Kossakowski matrix, evaluated 
at equilibration time,  $\mathbf{A}(\infty)$ (see SM of Ref.~\cite{Guarnieri:2018}), which indicates that the master equation is not a well-defined quantum-mechanical evolution for every initial state and may lead to unphysical results~\cite{Benatti:2005,Rivas:2012}. 
In Fig.~\ref{fig:figure} we compare the SSC with the negativity of the Kossakowski matrix $\mathcal{N}_K \equiv \sum_i (|\mu_i|-\mu_i)/2$, where $\mu_i$ are the eigenvalues of $\mathbf{A}(\infty)$, as a function of the parameter $\lambda$ controlling the weak-coupling expansion ($\lambda \sim \epsilon^2$, where $\epsilon \ll \omega_0$ is the system-bath coupling strengh).
Since $\Delta_1$, $\Delta_2$ and $J_\mathrm{eff}$ are all proportional to $\lambda$, it follows from Eq.~\eqref{eq:v1} that $\mathcal{C} \propto \lambda$ when $\lambda \rightarrow 0$.  
Remarkably, also $\mathcal{N}_K \propto \lambda$, and therefore non-zero coherence in the energy basis is in one to one correspondence to the loss of complete positivity. 
In addition, negativities in the steady-state density operator $\rho_\mathrm{ss}$ are obtained at low temperatures (see inset). 
Applying the standard Davies' (weak-coupling) limit~\cite{Davies:1974} by performing the secular approximation (see, e.g., Eqs.~(120) and~(123) of Ref.~\cite{Benatti:2005}), cross-terms of the form $\sigma_z \rho \sigma_\pm$, which rotate fast as $e^{\mp i\omega_0 t}$, disappear. This 
leads to two independent dissipators in Lindblad form, predicting a thermal steady state without SSC.

Summarizing, the high values for SSC generation reported in Ref.~\cite{Guarnieri:2018} for weak couplings (Eqs.(2)-(4) and Fig. 2) are erroneous, and where SSC survive, it is small and related to negativity issues at the order of $\lambda \sim \epsilon^2$.
The latter are cured by the application of the usual Davies' limit, not allowing for SSC. On the other hand, the corrected Eq.~\eqref{eq:v1}, after a truncation of $O(\lambda^2)$-terms in its regime of validity, reproduces the $\epsilon^2$-order result from the global thermal state perturbation expansion obtained in the SM, and is consistent with Ref.~\cite{Purkayastha:2019}. These results provide a small perturbative correction to the zero SSC value, whose full consistency would need to be checked at the fourth-order master equation level~\cite{Fleming:2011,Subas2012}, and hence should still be taken with a grain of salt. This point is specially important to ensure consistency with the second law of thermodynamics. We remark that standard quantum thermodynamics in the weak-coupling regime~\cite{Alicki:1979} (where the system-bath interaction Hamiltonian can be neglected) is only recovered in the Davies' limit. Otherwise strong-coupling-like corrections~\cite{Perarnau-Llobet:2018}, e.g. the explicit decoupling of the system from the bath~\cite{Hovhannisyan:2020}, must be taken into account to 
compensate for the work-value of the generated coherence~\cite{Skrzypczyk:2014,Korzekwa:2016,Manzano:2019}, and to comply with 
no-coherence-broadcasting results~\cite{Muller:2019,Spekkens:2019}. This implies considering extra (higher-order) corrections to the required resources that may spoil the autonomy claimed in Ref.~\cite{Guarnieri:2018}.

\begin{acknowledgments}
We thank the authors of Ref.~\cite{Guarnieri:2018}, and in particular G. Guarnieri, for their kind response and for useful insights that helped us to improve the presentation of this Comment. We also acknowledge R. Zambrini for interesting discussions and useful comments. M. Cattaneo is funded by the University of Helsinki and by the Severo Ochoa and María de Maeztu Program for Centers and Units of Excellence in R\&D (MDM-2017-0711). G. Manzano acknowledges financial support from Spanish MICINN through the program Juan de la Cierva-Incorporaci\'on  (IJC2019-039592-I).
\end{acknowledgments}

\bibliography{refs}

\end{document}